\documentclass[11pt]{article}
\usepackage{amsmath,amssymb}

\def\xxx#1           {{\sf hep-th/#1} }

\def\npb#1(#2)#3     {Nucl. Phys. {\bf B#1} (#2) #3 }
\def\rep#1(#2)#3     {Phys. Rept. {\bf #1} (#2) #3 }
\def\pla#1(#2)#3     {Phys. Lett. {\bf A#1} (#2) #3 }
\def\plb#1(#2)#3     {Phys. Lett. {\bf B#1} (#2) #3 }
\def\prl#1(#2)#3     {Phys. Rev. Lett. {\bf #1} (#2) #3 }
\def\prd#1(#2)#3     {Phys. Rev. {\bf D#1} (#2) #3 }
\def\ap#1(#2)#3      {Ann. Phys. {\bf #1} (#2) #3 }
\def\rmp#1(#2)#3     {Rev. Mod. Phys. {\bf #1} (#2) #3 }
\def\cmp#1(#2)#3     {Comm. Math. Phys. {\bf #1} (#2) #3 }
\def\mpla#1(#2)#3    {Mod. Phys. Lett. {\bf A#1} (#2) #3 }
\def\ijmp#1(#2)#3    {Int. J. Mod. Phys. {\bf A#1} (#2) #3 }
\def\cqg#1(#2)#3     {Class. Quant. Grav. {\bf #1} (#2) #3 }
\def\am#1(#2)#3      {Adv. Math. {\bf #1} (#2) #3 }
\def\im#1(#2)#3      {Invent. Math. {\bf #1} (#2) #3 }
\def\jhep#1(#2)#3    {JHEP {\bf #1} (#2) #3 }
\def\npps#1(#2)#3    {Nucl. Phys. Proc. Suppl. {\bf #1} (#2) #3 }
\def\jgp#1(#2)#3     {J. Geom. Phys. {\bf #1} (#2) #3 }
\def\atmp#1(#2)#3    {Adv. Theor. Math. Phys. {\bf #1} (#2) #3}
\def\lmp#1(#2)#3     {Lett. Math. Phys. {\bf #1} (#2) #3}
\begin{document}
\thispagestyle{empty}
\null\vskip-24pt \hfill LAPTH-857/01

\begin{center}
\vskip 0.2truecm {\Large\bf
On the OPE of 1/2 BPS Short Operators in $N=4$ SCFT$_4$}\\
\vskip 0.5truecm
{B. Eden\footnote{email:{\tt burkhard@lapp.in2p3.fr}} and
E. Sokatchev \footnote{email:{\tt sokatche@lapp.in2p3.fr} \\
}\\
\vskip 0.4truecm
{\it Laboratoire d'Annecy-le-Vieux de Physique
Th{\'e}orique\footnote{UMR 5108 associ{\'e}e {\`a}
l'Universit{\'e} de Savoie} LAPTH, Chemin de Bellevue, B.P. 110,
F-74941 Annecy-le-Vieux, France}}
\end{center}

\vskip 1truecm \Large
\centerline{\bf Abstract} \normalsize The content of the OPE of
two 1/2 BPS operators in $N=4$ SCFT$_4$ is given by their
superspace three-point functions with a third, a priori long
operator. For certain 1/2 BPS short superfields these three-point
functions are uniquely determined by superconformal invariance. We
focus on the cases where the leading ($\theta=0$) components lie
in the tensor products
$[0,m,0]\otimes[0,n,0]$ and  $[m,0,0]\otimes[0,0,n]$ of $SU(4)$.\\
We show that the shortness conditions at the first two points
imply selection rules for the supermultiplet at the third point.
Our main result is the identification of all possible protected
operators in such OPE's. Among them we find not only BPS short
multiplets, but also series of special long multiplets which
satisfy current-like conservation conditions in superspace.

\newpage
\setcounter{page}{1}\setcounter{footnote}{0}

\section{OPE, three-point functions and 1/2 BPS short superfields}\label{OPE}

The OPE of two conformal primary operators ${\cal \phi}(1)$ and
${\cal \psi}(2)$ can be written symbolically in the following form
\cite{FGG,DPPT,Pisa,FP}:
\begin{equation}\label{1}
  {\cal \phi}(1){\cal \psi}(2) =
  \sum_{(\sigma)}\int_{3,3'} \langle {\cal \phi}(1){\cal \psi}(2){\cal
  O}_{(\sigma)}(3)\rangle\ \langle {\cal O}_{(\sigma)}(3){\cal
  O}_{(\sigma)}(3')\rangle^{-1}\ {\cal O}_{(\sigma)}(3')\,.
\end{equation}
Thus, the OPE has the meaning of the tensor product of two UIR's
of the four-dimensional conformal group $SO(1,5)$ (Euclidean case)
or $SO(2,4)$ (Minkowski case) decomposed into an infinite
(discrete or continuous) sum of UIR's denoted by ${\cal
O}_{(\sigma)}$. In this decomposition the r\^ole of the
Clebsch-Gordon coefficients (the kernel of the integral
decomposition (\ref{1})) is played by the three-point function of
the two operators with each operator appearing in the OPE,
multiplied (in the sense of convolution) by the inverse two-point
function of the latter.\footnote{The more familiar form of the OPE
${\cal \phi}{\cal \psi} = \sum C(x,\partial_x){\cal O}$ is
obtained by Taylor-expanding the three-point function in the limit
$1\rightarrow2$ \cite{FP}.}

It should be pointed out that there is a subtle but very important
difference between the Euclidean and Minkowski cases
\cite{FGPa,DPPT}. In the Euclidean case the inverse two-point
function is simply obtained by changing the dimension of the
operator (``shadow" operator), but the sum $\sum_{(\sigma)}$
should be understood as an integral over the imaginary conformal
dimension. In the Minkowski case the spectrum of the dimension
becomes discrete, but the inverse two-point function is a
non-trivial object which makes the kernel in (\ref{1}) transform
non-locally. Nevertheless, in both cases the content of the OPE
can be determined by examining the three-point functions that the
two operators can possibly form with a third, a priori arbitrary
operator.

We shall assume that the above applies to the $N=4$ superconformal
algebra $PU(2,2/4)$,\footnote{The R charge of the $N=4$
superconformal algebra is a central charge. In $N=4$ SYM this
charge vanishes, therefore one usually considers the superalgebra
$PSU(2,2/4)$ without central charge. However, we prefer to work
with $PU(2,2/4)$ because we want to study the OPE of more general
superconformal operators, some of which may carry central charge.
We are grateful to S. Ferrara for turning our attention to this
point.} although to our knowledge no rigorous mathematical
analysis exists in the literature. So, we propose to reduce the
problem of studying the OPE of two superconformal primary
operators to constructing all possible three-point functions of
these operators with a third operator. In order to make conformal
supersymmetry manifest, it is natural to carry out the analysis in
superspace.\footnote{OPE's in SCFT have mostly been investigated
in ordinary space in terms of the components of the
supermultiplets \cite{Anselmi:1997mq}. In this way it is not easy
to fully exploit the power of conformal supersymmetry. An early
attempt on a superspace OPE was made in \cite{HWOPE}.} At first
sight, this task may seem difficult for the following reason. A
superspace three-point function depends on three sets of Grassmann
coordinates $\theta^\alpha_i$, $\bar\theta^{\dot\alpha i}$,
$i=1,\ldots,4$, i.e., on $3\times 4\times 4 = 48$ odd variables.
On the other hand, the superconformal algebra contains two odd
generators ($Q$ of Poincar\'e and $S$ of conformal supersymmetry)
acting on the $\theta$'s as $2\times 4\times 4 = 32$ shifts. Thus,
one can form $48-32=16$ invariant combinations of the Grassmann
coordinates. This implies that when expanding a super-three-point
function in the $\theta$'s, at each step new arbitrary nilpotent
terms may appear, up to the level $\theta^{16}$, thus allowing for
an enormous arbitrariness.

The situation radically changes if one puts certain types of 1/2
BPS short operators at points 1 and 2, still keeping the third
operator arbitrary. By definition, an $N$-extended 1/2 BPS short
superfield depends only on half of the Grassmann variables,
namely, on $p$ left-handed $\theta^\alpha$'s and on $N-p$
right-handed $\bar\theta^{\dot\alpha}$'s. Using the terminology of
\cite {hh} we can call them $(N|N-p,p)$ superfields. Let us put a
$(4|4-p_1,p_1)$ and a $(4|4-p_2,p_2)$ superfield at points 1 and
2, respectively. Repeating the above counting, we find $(p_1 + p_2
+ 4)\times 4$ left-handed and $(4-p_1 + 4-p_2 + 4)\times 4$
right-handed odd coordinates in the three-point function. If we
choose $p_1$ and $p_2$ such that $p_1+p_2=4$, then the above
numbers exactly match the numbers of left- and right-handed
supersymmetries. So, in this case there are no nilpotent
invariants and {\it the three-point superfunction is uniquely
determined by its leading component} (obtained by setting all
$\theta=0$). Our aim here will be to construct such three-point
functions and study the restrictions on the operators at point 3.

The 1/2 BPS short $N=4$ superfields fall into two essentially different
classes. One of them, familiar from $N=1$ supersymmetry, are the so-called
chiral superfields\footnote{The term ``chiral primary operator" is often
improperly used in the literature to denote all kinds of BPS short objects.},
i.e. $(4|4,0)$ or $(4|0,4)$ superfields. It should be stressed that the
conformal dimension of a chiral superfield is only limited by the unitarity
bound, so such objects do not have a protected dimension in SCFT. According to
our counting above, the three-point functions of a chiral and an antichiral
superfield with a third, general superfield are uniquely determined by
conformal supersymmetry. Three-point functions involving only (anti)chiral
superfields have been studied in \cite{Pickering:2000rk,DolOsb}, but no
detailed explicit results are available for the case relevant to the OPE where
the third superfield is general. A preliminary examination of this case does
not indicate any interesting selection rules on the superconformal UIR's in
such OPE's.

Extended ($N>1$) supersymmetry provides a new class of 1/2 BPS
short superfields called Grassmann (G-)analytic (they were
introduced for the first time in the $N=2$ case in \cite{GIO}).
Examples of such G-analytic objects are the $N=2$ matter multiplet
(hypermultiplet) \cite{GIK1} and the $N=3$ \cite{GIK2} and $N=4$
super-Yang-Mills multiplet \cite{Bandos,hh,hw1}. The essential
difference from the chiral case is that G-analytic superfields
depend on at least one $\theta$ and one $\bar\theta$. Moreover,
they carry no spin and their conformal dimension is expressed in
terms of the Dynkin labels of the R symmetry representation of
their first component, i.e., they have a {\it protected dimension}
in  SCFT \cite{dp}. Such objects have an important r\^ole in the
AdS/CFT correspondence because they are the SCFT duals of the KK
states in the AdS bulk \cite{Gunaydin:1985fk,AF}.

The G-analytic superfields are of the type $(4|4-p,p)$ with
$p\neq0,4$. With the additional requirement that the three-point
function of two such operators should be uniquely determined by
conformal supersymmetry, our choice for the two 1/2 BPS operators
whose OPE we want to study is restricted to: (i) a pair of
self-conjugate superfields $(4|2,2)$; (ii) one $(4|1,3)$ and one
$(4|3,1)$ superfield. The OPE content of such operators is the
subject of the present paper. Here we develop the idea put forward
in the recent publication \cite{Arutyunov:2001qw} and apply it to
the case of $N=4$ supersymmetry. Our main result is the
identification of all possible protected operators in this OPE.
Among them we find not only BPS short multiplets, but also series
of special long multiplets which satisfy current-like conservation
conditions in superspace. One example of the latter is the
multiplet starting with a scalar of dimension 4 in the {\bf 20} of
$SU(4)$ whose ``exceptional" non-renormalization properties have
been discovered in \cite{AFP1} and later discussed in
\cite{AFP2,AEPS,BKRSkonishi}.

The paper is organized as follows. In Section 2 we recall the key points of the
harmonic superspace \cite{GIK1,TheBook} treatment of BPS short (or G-analytic)
superfields. We also give some basic information about the UIR's of the
superconformal algebra $PU(2,2/4)$ and identify the types of 1/2 BPS short
objects whose OPE we wish to discuss. Section 3 is devoted to the construction
of the three-point functions of these 1/2 BPS operators with an arbitrary third
operator.\footnote{Superconformal two- and three-point functions in superspace
have been discussed in many papers. These include the case of $N=1$ chiral
superfields \cite{Pickering:2000rk,DolOsb} as well as a method valid for all
types of $N=1$ \cite{Park,Osborn} and $N>1$ superfields \cite{Park2}. This
method has recently been adapted to G-analytic superfields in \cite{hw9}. A
construction of analytic $N=2$ and $N=4$ two- and three-point functions is
presented in \cite{hw1,EHW} (see also \cite{HoSoWe}). Ref. \cite{KuzTheis}
discusses the $N=2$ case in the context of three protected (current or
stress-tensor) operators. However, no explicit results are available for the
case of interest to us, namely, when two of the operators are short but the
third one is generic.} Here we do not present the complete form of these
functions. Instead, we use conformal supersymmetry to reconstruct the
dependence on the Grassmann coordinates at one short point only. This is
sufficient to examine the consequences of the shortness conditions. In this way
we derive selection rules for the superconformal UIR's at the third point. The
results are summarized in Section 4. There we also comment on the application
of our results to the four-point function of $N=4$ SYM stress-tensor multiplets
and to the so-called ``extremal" \cite{DHoFrMaMaRa} and ``next-to-extremal"
\cite{EdHoScSoWe} correlators.

\section{Grassmann-analytic superfields}

\subsection{On-shell $N=4$ SYM and Grassmann analyticity}

The simplest example of an $N=4$ G-analytic superfield is the SYM
field strength on shell. In ordinary superspace it is described by
a real scalar superfield $W^{ij}(x,\theta,\bar\theta)= -W^{ji} =
1/2 \epsilon^{ijkl}\bar W_{kl}$ in the {\bf 6} of $SU(4)$ (Dynkin
labels $[0,1,0]$) satisfying the following on-shell constraints
(massless field equations):
\begin{equation}\label{1-1}
  D^{(k}_\alpha W^{i)j}=0\;,\qquad \bar D_{\dot\alpha \{k}W^{i\}j}=0
\end{equation}
where $()$ means symmetrization, $\{\}$ denotes the traceless part
and $D^{k}_\alpha$, $\bar D_{\dot\alpha\;k}$ are the superspace
covariant derivatives obeying the algebra
\begin{equation}\label{14}
  \{D^i_\alpha,D^j_\beta\} = \{\bar D_{\dot\alpha i},\bar D_{\dot\beta j}\} =
  0\;, \quad \{D^i_\alpha,\bar D_{\dot\beta j}\} = -2i\delta^i_j
  (\sigma^\mu)_{\alpha\dot\beta}\partial_\mu\;.
\end{equation}

These constraints can be equivalently rewritten in the form of
G-analyticity conditions without losing the manifest $SU(4)$
invariance with the help of the so-called harmonic variables
(first introduced in \cite{GIK1} in the case $N=2$ and then
generalized to $N>2$ in \cite{GIK2,Bandos,hh}). The harmonics
$u^I_i$ (and their conjugates $u^i_I = (u^I_i)^* $) form an
${SU}(4)$ matrix where $i$ is an index in the fundamental
representation of ${SU}(4)$ and $I=1,\ldots,4$ are the projections
of the second index onto the subgroup $[{U}(1)]^{3}$. Further, we
define two {\sl independent} ${SU}(4)$ groups, a left one acting
on the index $i$ and a right one acting on the projected index $I$
of the harmonics:
\begin{equation}\label{3.0}
 (u^I_i)'= \Lambda_i^ju_j^J\Sigma_J^I\;, \qquad \Lambda\in{SU}(4)_L\;, \quad
\Sigma\in{SU}(4)_R\;.
\end{equation}
In particular, the charge operators (the three generators of the
Cartan subgroup $[{U}(1)]^{3}\subset {SU}(4)_R$) act on the
harmonics as follows:
\begin{equation}\label{3.1}
m_K\; u^I_i = (\delta^I_{K}-\delta^I_{4}) u^I_i\;, \qquad m_K\; u^i_I = -
(\delta_{KI}-\delta_{4I})  u^i_I\;,
\end{equation}
so that $m_4\equiv 0$. The harmonics satisfy the following ${SU}(4)$ defining
conditions:
\begin{eqnarray}
 &&u^I_i u^i_J=\delta^I_J~, \nonumber\\ u\in
{SU}(4):\quad &&u^I_i u^j_I =\delta^j_i~,\label{3.2}\\ &&
\varepsilon^{i_1\ldots i_4}u^1_{i_1}\ldots u^4_{i_4}=1~. \nonumber
\end{eqnarray}

Now, let us use the harmonic variables to covariantly project all
the $SU(4)$ indices in the constraints (\ref{1-1}) onto
$[\mbox{U}(1)_R]^{3}$. In this way all objects become $SU(4)_L$
invariant but $[\mbox{U}(1)_R]^{3}$ covariant. For example, the
projection
\begin{equation}\label{4.12}
  W^{12} = W^{ij}(x,\theta,\bar\theta) u^1_{i}u^2_{j}
\end{equation}
satisfies the constraints
\begin{equation}\label{4.14}
  D^{1}_\alpha  W^{12} = D^{2}_\alpha  W^{12} =
   \bar D_{\dot\alpha\; 3} W^{12} = \bar D_{\dot\alpha\; 4} W^{12} = 0
\end{equation}
where
\begin{equation}\label{3.8}
D^I_\alpha = D^i_\alpha u_i^I\;, \quad \bar D^{\dot\alpha}_J = \bar
D^{\dot\alpha}_i u^i_J\;.
\end{equation}
{}From (\ref{14}), (\ref{3.2}) and (\ref{3.8}) it follows that the projected
derivatives appearing in (\ref{4.14}) anticommute. This in turn implies the
existence of a G-analytic basis in superspace,
\begin{eqnarray}
  &&x^\mu_A =  x^\mu
-2i(\theta_1\sigma^\mu \bar\theta^{1} +  \theta_2 \sigma^\mu \bar\theta^{2} -
\theta_{3} \sigma^\mu \bar\theta^{3}- \theta_4 \sigma^\mu \bar\theta^{4}
)\;,\nonumber\\
  &&\theta^\alpha_I = \theta^\alpha_i u^i_I\;, \quad
\bar\theta^{\dot\alpha I} = \bar\theta^{\dot\alpha i} u_i^I\;,\label{3.9}
\end{eqnarray}
where these derivatives become just $D^{1,2}_\alpha =
\partial/\partial\theta^\alpha_{1,2}$, $\bar D_{\dot\alpha\; 3,4} =
-\partial/\partial\bar\theta^{\dot\alpha\; 3,4}$. Consequently, in this basis
the superfield $W^{12}$ is an unconstrained {\it Grassmann-analytic} function
of two $\theta$'s and two $\bar\theta$'s, as well as of the harmonic variables:
\begin{equation}\label{4.16}
   W^{12} =  W^{12}(x_A,\theta_{3},\theta_{4}, \bar\theta^1,\bar\theta^2, u)\;.
\end{equation}
So, this is an example of a 1/2 BPS short superfield of the type $(4|2,2)$.

\subsection{$SU(4)$ irreducibility and harmonic analyticity}

It is important to realize that the G-analytic superfield
(\ref{4.16}) is an $SU(4)_L$ invariant object only because it
is a function of the harmonic variables. In order to recover the
original harmonic-independent but constrained superfield
$W^{ij}(x,\theta,\bar\theta)$ (\ref{1-1}) we need to impose
harmonic differential constraints equivalent to the definition of
a highest weight state (HWS) of $SU(4)$. To this end we introduce
derivatives acting on the harmonics as follows:
\begin{equation}
\partial^I_J u^K_i=\delta^K_J u^I_i - \frac{1}{4}\delta^{I}_J u^K_i \;,\qquad
\partial^I_J u^i_K=-\delta^I_K u^i_J +  \frac{1}{4}\delta^{I}_J u^i_K\; .\label{3.5}
\end{equation}
It is easy to see that they generate the group ${SU}(4)_R$ acting
on the $[{U}(1)_R]^{3}$ projected indices of the harmonics.

A harmonic function defined on $SU(4)$ can be restricted to the
coset $SU(4)/[U(1)]^3$ by assuming that it transforms
homogeneously under $[{U}(1)_R]^{3}\subset SU(4)_R$. In terms of
the harmonic derivatives this condition is translated into the
requirement that the harmonic functions $f(u)$ are eigenfunctions
of the diagonal derivatives $\partial^{\,I}_I-
\partial^{\,4}_4$ which count the ${U}(1)_R$ charges (\ref{3.1}).
Then the independent harmonic derivatives on the coset are the six complex
derivatives $\partial^{\,I}_J $, $I<J$ corresponding to the raising operators
in the Cartan decomposition of the algebra of ${SU}(4)_R$.

In general, the harmonic functions have an infinite ``harmonic" expansion on
the coset\linebreak $SU(4)/[U(1)]^3$, thus giving rise to an infinite set of
$SU(4)$ irreps. They can be restricted to a single irrep by requiring that they
are annihilated by the raising operators of ${SU}(4)_R$, i.e. that they
correspond to a HWS,
\begin{equation}\label{a10}
  \partial^{\,I}_J f(u) = 0\;, \quad I<J\,.
\end{equation}
For example, the harmonic function $f^1(u)$ subject to the constraint
(\ref{a10}) is reduced to the {\bf 4}: $f^i u^1_i$; the function $f^{12}(u)$ to
the {\bf 6}: $f^{ij}u^1_i u^2_j$ where $f^{ij}=-f^{ji}$; the function
$f^{123}(u)\equiv \bar f_4(u)$ to the {$\bar{\mbox{\bf 4}}$}: $f^{ijk}u^1_i
u^2_j u^3_k = \bar f_i u^i_4$ where $f^{ijk} = \epsilon^{ijkl}\bar f_l$, etc.
The generalization is straightforward: the harmonic function
\begin{equation}\label{3.333}
  f^{\stackrel{\underbrace{\mbox{\scriptsize
1\ldots 1}}}{m_1} \stackrel{\underbrace{\mbox{\scriptsize 2\ldots 2}}}{m_2}
\stackrel{\underbrace{\mbox{\scriptsize 3\ldots 3}}}{m_{3}}}(u)\;, \qquad { m_1
\geq m_2\geq  m_{3}\geq 0 }
\end{equation}
subject to the irreducibility condition (\ref{a10}) corresponds to the HWS of
the $SU(4)$ irrep carrying the Young tableau labels $m_1,m_2,m_3$ (recall
(\ref{3.1})) or the Dynkin labels $a_1=m_1-m_2,\ a_2=m_2-m_3,\ a_3=m_3$.

The $SU(4)$ irreducibility conditions (\ref{a10}) have an
alternative interpretation obtained by introducing complex
coordinates $z_m$, $m=1,\ldots,6$ on the harmonic coset
$SU(4)/[U(1)]^3$. Then the six derivatives in (\ref{a10})
correspond, roughly speaking, to $\partial/\partial \bar z_m$. So,
conditions (\ref{a10}) are (covariant) Cauchy-Riemann analyticity
conditions on the compact complex manifold $SU(4)/[U(1)]^3$. The
only {\it regular} solutions allowed by such constraints are
homogeneous harmonic polynomials, i.e., $SU(4)$ irreps. In this
sense one may call (\ref{a10}) harmonic (H-)analyticity conditions
\cite{hh} .

It is important to realize that there also exist {\it singular}
harmonic functions which do not correspond to $SU(4)$ irreps. To
give an example, consider two sets of harmonics, $u^I_i$ and
$v^i_I$, and form the $SU(4)_L$ invariant combination, e.g.,
$u^1_iv^i_2$. Obviously, it is H-analytic, i.e.,
$(\partial^{\,I}_J)_u u^1_iv^i_2 = 0$, $I<J$. However, this is not
true for the function $(u^1_iv^i_2)^{-1}$ which becomes singular
when $u=v$ (recall (\ref{3.2})). In fact, the raising operators
acting on this singular function produce harmonic delta functions,
in accordance with the formula $\partial/\partial \bar z_m
z^{-1}_m = \pi \delta(z_m)$. In other words, the singular harmonic
functions cannot be expanded on the coset  $SU(4)/[U(1)]^3$ into a
harmonic series of $SU(4)$ irreps. The issue of harmonic
singularities \cite{hw1,Pickering} will be the central point in
our study of the three-point functions in Section 3.

Let us now come back to the $N=4$ SYM field-strength in the form of the
G-analytic superfield (\ref{4.16}). The analog of (\ref{a10}) in this case are
the H-analyticity conditions
\begin{equation}
  \label{a10'}
  D^{\,I}_J W^{12} = 0\;, \qquad  I<J
\end{equation}
where $D^{\,1}_2 = \partial^{\,1}_2$, $D^{\,3}_4 =
\partial^{\,3}_4$ and
\begin{equation}\label{covdev}
  D^{\,I}_J=
\partial^{\,I}_{J} +2 i\theta_{J}\sigma^\mu
\bar\theta^{I}\partial_\mu - \theta_J\partial^I + \bar\theta^I\bar\partial_J\,,
\qquad I=1,2, \ J=3,4
\end{equation}
are the supercovariant versions of the harmonic derivatives in the G-analytic
basis (\ref{3.9}). The presence of space-time derivatives in (\ref{covdev})
implies massless field equations on the component fields, which are the six
physical scalars, the four spinors and the vector of the $N=4$ SYM on-shell
multiplet.

We remark that the conditions of G-analyticity and H-analyticity are compatible
because the spinor derivatives from  (\ref{4.14}) commute with the harmonic
ones from  (\ref{a10'}). One says that they form a Cauchy-Riemann (CR)
structure \cite{Rosly}.

\subsection{Other types of G- and H-analytic objects}

The $N=4$ SYM field-strength is the simplest example of a G- and H-analytic
$N=4$ superfield. Before considering its generalizations, we remark that the
projection (\ref{4.12}) is not the only possible way to convert $W^{ij}$ into a
G-analytic superfield. An equivalent projection is
\begin{equation}\label{4.120}
  W^{13} = W^{ij}(x,\theta,\bar\theta) u^1_{i}u^3_{j}
\end{equation}
and the resulting G-analytic superfield depends on a different half of the
Grassmann variables,
\begin{equation}\label{4.160}
    W^{13}(x_A,\theta_{2},\theta_{4}, \bar\theta^1,\bar\theta^3, u)\;.
\end{equation}
According to our conventions, this projection is not a HWS of an $SU(4)$ irrep,
so one of the raising operators converts it into the HWS $W^{12}$:
\begin{equation}\label{4.161}
  D^{\,2}_{3} W^{13}(\theta_{2,4}, \bar\theta^{1,3}) =
  W^{12}(\theta_{3,4}, \bar\theta^{1,2})\,,
\end{equation}
whereas all the remaining raising operators still annihilate it.

Besides the self-conjugate $N=4$ SYM multiplet, there exist two
other massless conjugate $N=4$ multiplets which have no spin but
carry non-trivial $SU(4)$ quantum numbers \cite{Siegel}. In
ordinary superspace they are described by a superfield
$W^i(x,\theta,\bar\theta)$ in the fundamental of $SU(4)$ and by
its conjugate $\bar W_i \equiv - \frac{1}{6} \, \epsilon_{ijkl} W^{jkl}$
satisfying on-shell constraints similar to (\ref{1-1}). In harmonic
superspace this corresponds to the G-analytic superfields
\begin{equation}\label{0.1}
  W^1(x_A, \theta_2,\theta_3,\theta_4,\bar\theta^1,u)\quad \mbox{or} \quad
   W^{123}(x_A, \theta_4,\bar\theta^1,\bar\theta^2,\bar\theta^3,u)\equiv \bar
   W_4
\end{equation}
satisfying the H-analyticity conditions $D^{\,I}_J W^1=0$, $D^{\,I}_J \bar
W_4=0$, $I<J$ (cf. (\ref{a10'})). Note that the G-analytic basis (\ref{3.9})
needs to be adapted to the new types of G-analyticity. We can say that $W^1$ is
1/2 BPS short of the type $(4|1,3)$ and its conjugate $\bar W_4$ of the type
$(4|3,1)$.

Further BPS short objects can be constructed by multiplying the
above ones. For instance, an interesting class of 1/2 BPS
superfields of the type $(4|2,2)$ is obtained as gauge-invariant
composite operators built out of the SYM field-strength
\cite{hw1}:
\begin{equation}\label{4.162}
  {\cal O}^{[0,m,0]}_{1/2} = {\rm Tr}\ [W^{12}(\theta_{3,4},
  \bar\theta^{1,2})]^m\,.
\end{equation}
Since the elementary superfield $W^{12}$ is subject to the irreducibility
constraints (\ref{a10'}), the same applies to the composite as well. We
conclude that its leading component is in the $SU(4)$ irrep with Dynkin labels
$[0,m,0]$. Further, the leading component of $W^{12}$ is a massless scalar of
canonical dimension $\ell=1$, therefore the composite object (\ref{4.162}) has
dimension $\ell = m$. If $m>1$ the composites (\ref{4.162}) are not massless.
The case $m=2$ corresponds to the $N=4$ SYM stress-tensor multiplet. Note that
the operators of this series are self-conjugate, just like the SYM
field-strength itself.

Similarly, using the alternative massless multiplets (\ref{0.1}), we obtain two
conjugate series:
\begin{equation}\label{0.2}
  {\cal O}^{[m,0,0]}_{1/2} =  [W^{1}(\theta_{2,3,4},
  \bar\theta^{1})]^m\quad \mbox{and} \quad {\cal O}^{[0,0,m]}_{1/2} =
   [\bar W_4(\theta_{4}, \bar\theta^{1,2,3})]^m\,.
\end{equation}
Note that we have dropped the YM trace in (\ref{0.2}), treating
the multiplication of $W$'s as a formal way of obtaining new
superconformal UIR's \cite{AFSZ,Ferrara:2000zg,Ferrara:2001eb}. As
before, these objects have dimension $\ell=m$.

Next, mixing different types of G-analytic objects we can obtain BPS objects
with a lower degree of shortening. For instance, the following 1/4 BPS
superfields:
\begin{equation}\label{4.163}
  {\cal O}^{[q,p,q]}_{1/4}  = [W^{12}(\theta_{3,4}, \bar\theta^{1,2})]^{p+q}
  \ [W^{13}(\theta_{2,4}, \bar\theta^{1,3})]^{q}\,, \qquad \ell=p+2q
\end{equation}
and
\begin{equation}\label{0.3}
  {\cal O}^{[p,0,q]}_{1/4}  = [W^{1}(\theta_{2,3,4}, \bar\theta^{1})]^{p}
  \ [\bar W_4(\theta_{4}, \bar\theta^{1,2,3})]^{q}\,, \qquad \ell=p+q
\end{equation}
are G-analytic of the type $(4|1,1)$. They appear in the OPE of 1/2 BPS
objects, as we show in the next section.

\subsection{Series of UIR's of $PU(2,2/4)$}

The UIR's of $PU(2,2/4)$ are labeled \cite{dp} by the conformal
dimension $\ell$, the two Lorentz ($SO(4)$ or $SO(1,3)$) quantum
numbers (``spins") $j_1,j_2$ and the $SU(4)$ Dynkin labels
$a_1,a_2,a_{3}$:
\begin{equation}\label{7}
  {\cal D}(\ell;j_1,j_2;a_1,a_2, a_{3})\,.
\end{equation}
In addition, in $PU(2,2/4)$ there is a central charge which does
not intervene in our discussion, so we do not list it among the UIR
labels. Above we saw examples of G-analytic BPS operators of the
type $(4|p,q)$, $pq\neq0$. According to the classification of
\cite{dp} (in the notation of
\cite{AFSZ,Ferrara:2000zg,Ferrara:2001eb}), they belong to the
series C of UIR's. The most general representative of this series
can be viewed as a product of the three kinds of massless
superfields $W^1$, $W^{12}$ and $W^{123}$ \cite{Ferrara:2001eb}:
\begin{equation}\label{seriesC}
  [W^1]^{a_1}[W^{12}]^{a_2}[W^{123}]^{a_3}\,.
\end{equation}
This object has no spin, $j_1=j_2=0$, but carries a non-trivial
$SU(4)$ irrep with Dynkin labels $[a_1,a_2,a_3]$. Moreover, its
conformal dimension is {\it fixed}:
\begin{equation}\label{Cdim}
  \ell = a_1+a_2+a_3\,.
\end{equation}
In a quantum theory this means that such operators cannot acquire
an anomalous dimension or, as one says, they are {\it protected}
by conformal supersymmetry.

The other class of BPS short multiplets form the series B of UIR's of
$PU(2,2/4)$. They are of the type $(4|0,p)$ (or $(4|p,0)$). The familiar chiral
(or antichiral) superfields correspond to $p=4$. This time they can carry both
spin $(j_1,0)$ (or $(0,j_2)$) and $SU(4)$ quantum numbers, and their dimension
is limited only by the unitarity bound, e.g., $\ell \geq 1+j_1+a_1+a_2+a_3$.
Consequently, operators from this series are in general not
protected.\footnote{Note that the situation changes in the case of the
superalgebra $PSU(2,2/4)$. There the R (or central) charge vanishes and the
dimension of the series B UIR's gets fixed, i.e., they become protected.}

Finally, the series A involves multiplets without any BPS
shortening. They carry arbitrary spin and $SU(4)$ quantum numbers.
In this case the unitarity bound is (upon elimination of the
central charge)
\begin{equation}\label{seriesA}
  \ell \geq 2+j_1+j_2+a_1+a_2+a_3\,.
\end{equation}
There is no reason for these operators to be protected unless the unitarity
bound is saturated. To explain this, recall the well-known example of a
conformal vector field $J_\mu(x)$. Its dimension should satisfy the unitarity
bound $\ell\geq 3$. In general, this field is irreducible as a representation
of the conformal group, but it becomes reducible (although indecomposable) when
the unitarity bound is saturated, $\ell=3$. In this case the divergence of the
vector $\partial^\mu J_\mu(x)$ transforms covariantly and can thus be set to
zero. The resulting transverse (conserved) vector is an exceptional UIR with
fixed (i.e., ``protected") dimension\footnote{ We call such objects
``protected'', although the mechanism keeping the current conserved
needs to be explained.}. In CFT this corresponds to a conserved
current operator made out of elementary massless physical fields, e.g., $J_\mu
= \bar\phi\partial_\mu \phi - \phi \partial_\mu \bar\phi$. Furthermore, the
three-point function $\langle\phi_d(x)\phi_d(y)J^\mu_\ell(z)\rangle$ of two
scalars and a vector of dimension $d$ and $\ell$, respectively, is completely
determined by conformal invariance. At the unitarity bound $\ell=3$ it is
automatically conserved, $\langle\phi_d(x)\phi_d(y)J^\mu_\ell(z)\rangle
\overleftarrow{\partial}_{z\;\mu}=0$.

Exactly the same phenomenon takes place in series A. When the unitarity bound
is saturated,
\begin{equation}\label{unitbound}
  \ell = 2+j_1+j_2+a_1+a_2+a_3\,,
\end{equation}
the operator $J^{[a_1,a_2,a_3]}_{\alpha_1\ldots\alpha_{2j_1};\
\dot\alpha_1\ldots\dot\alpha_{2j_2}}$ realizes a reducible representation of
$PU(2,2/4)$. As in the case of the current, we can impose the ``conservation"
conditions:
\begin{equation}\label{currcon1}
  D^{1\;\alpha}\, J^{[a_1,a_2,a_3]}_{\alpha\alpha_2\ldots\alpha_{2j_1};\
   \dot\alpha_1\ldots\dot\alpha_{2j_2}} =
  \bar D_4^{\dot\alpha}\, J^{[a_1,a_2,a_3]}_{\alpha_1\ldots\alpha_{2j_1};\
   \dot\alpha\dot\alpha_2\ldots\dot\alpha_{2j_2}}
  =0
\end{equation}
if $s\neq0$ or
\begin{equation}\label{currcon2}
  D^{1\;\alpha}D^1_\alpha\, J^{[a_1,a_2,a_3]} = \bar D_{4\;\dot\alpha} \bar D_4 ^{\dot\alpha}\,
  J^{[a_1,a_2,a_3]} = 0
\end{equation}
if $s=0$, or similar ones involving more projections of the spinor derivatives
for some restricted sets of Dynkin labels \cite{Ferrara:2001eb}. In this sense
we may call objects of the type (\ref{currcon1}), (\ref{currcon2})
``current-like". It should be stressed that in general none of the components
of such an object is a conserved tensor. This only happens when $J$ is an
$SU(4)$ singlet. In this case the complete spinor derivatives appear in
(\ref{currcon1}), (\ref{currcon2}) and $J$ can be viewed as a bilinear
composite made out of massless chiral superfields, i.e., as a conserved
generalized supercurrent.

The differential constraints (\ref{currcon1}), (\ref{currcon2}) ``protect" the
dimension of such objects, just like the conservation condition $\partial^\mu
J_\mu=0$ protects the dimension of the current. In Section 3 we show that they
can appear in the OPE of two 1/2 BPS operators.

\section{Three-point functions involving two 1/2 BPS operators}

The main aim of the present paper is to study the OPE of two 1/2 BPS operators
of the type ${\cal O}^{[0,m,0]}_{1/2}\; {\cal O}^{[0,n,0]}_{1/2}$ or ${\cal
O}^{[m,0,0]}_{1/2}\; {\cal O}^{[0,0,n]}_{1/2}$. To this end we have to
construct all possible superconformal three-point functions
\begin{equation}\label{5.1}
  \langle {\cal O}^{[0,m,0]}_{1/2}(1)
  {\cal O}^{[0,n,0]}_{1/2}(2)
  {\cal O}^{{\cal D}}(3)\rangle
\end{equation}
and
\begin{equation}\label{5.1'}
  \langle {\cal O}^{[m,0,0]}_{1/2}(1)
  {\cal O}^{[0,0,n]}_{1/2}(2)
  {\cal O}^{{\cal D}}(3)\rangle
\end{equation}
where ${\cal O}^{{\cal D}}$ is an a priory arbitrary operator
carrying the $PU(2,2/4)$ UIR ${\cal D}$ (\ref{7}). Our task is to
find out what the allowed UIR's are. As explained in Section
\ref{OPE}, conformal supersymmetry completely determines such
three-point functions. The restrictions on the third UIR, apart
from the standard conformal ones, stem from the requirement of
H-analyticity ($SU(4)$ irreducibility) at points 1 and 2 combined
with G-analyticity. As we shall see later on, the expansion of
(\ref{5.1}) and (\ref{5.1'}) in, e.g., the $\theta$'s at point 1
may contain harmonic singularities.\footnote{The full $\theta$
dependence can be obtained by adapting the standard method of
\cite{Park,Osborn} and will be given elsewhere. It is
rather complicated and is irrelevant to the issue of harmonic
singularities at points 1 or 2.} If this happened, the basic
assumption that we are dealing with $SU(4)$ UIR's would not be
true. To put it differently, harmonic singularities violate
H-analyticity, i.e., $SU(4)$ irreducibility. Demanding that such
singularities be absent implies selection rules for the UIR's at
point 3.

Although the case (\ref{5.1}) is the physically interesting one (the operators
${\cal O}^{[0,m,0]}_{1/2}$ can be viewed as composites made out of the $N=4$
SYM field-strength, see (\ref{4.162})), the second case (\ref{5.1'}) is simpler
and we present it in more detail to explain our method. The generalization to
the case (\ref{5.1}) is straightforward.

\subsection{Two-point functions in the case ${\cal O}^{[m,0,0]}_{1/2}
  {\cal O}^{[0,0,n]}_{1/2}$}

Before discussing the three-point function (\ref{5.1'}) itself, it is
instructive to examine the two-point function
\begin{equation}\label{5.101}
  \langle W^{1}(x,\theta_{2,3,4}, \bar\theta^{1},1)
  \bar W_4(y, \zeta_4, \bar\zeta^{1,2,3},2) \rangle
\end{equation}
where the two sets of harmonic variables are denoted by $1^I_i$ and $2^I_i$,
respectively. Its leading component (obtained by setting $\theta=\zeta=0$) is
easily found:
\begin{equation}\label{5.3}
  \langle W^{1}\bar W_4\rangle_{\theta=\zeta=0} = \frac{(1^{1}2_4)}{
  (x-y)^2}
\end{equation}
where \begin{equation}\label{5.4}
  (1^1 2_4) \equiv 1^1_i 2^i_4
\end{equation}
is the $SU(4)_L$ invariant contraction of the harmonics at points
1 and 2. {}From (\ref{3.5}) it is clear that it satisfies the
$SU(4)$ irreducibility (H-analyticity) condition (\ref{a10}) at
point 1,
\begin{equation}\label{5.5}
  (\partial^I_J)_1\, (1^1 2_4) = 0\,, \quad I<J
\end{equation}
and similarly at point 2. The space-time factor in (\ref{5.3}) is determined by
the other quantum numbers of $W$, namely, spin 0 and dimension 1.

The next question is how to restore the odd coordinate dependence
starting from (\ref{5.3}). The counting argument of Section
\ref{OPE} shows that the two-point function (\ref{5.101}) is
overdetermined since it only depends on $16$ odd variables,
compared to the $32$ supersymmetries. Thus, it is sufficient to
use only $Q$ supersymmetry to fully restore the odd coordinates.
Furthermore, our aim is to study the harmonic singularities at
point 1 (or, equivalently, at point 2), so it is enough to restore
the $\theta$ dependence. To see this, imagine a frame in which
$\zeta$ has been set to zero by a finite $Q$-supersymmetry
transformation. Such a transformation depends on the harmonic
$2^I_i$ only, so it cannot introduce singularities at point 1. In
this frame the residual $Q$ supersymmetry is given by the
conditions (recall (\ref{3.2}))
\begin{eqnarray}
  \delta'_Q \zeta_4 = 0 &\Rightarrow& \epsilon'_i =
  (2^1_i 2_1^j + 2^2_i 2_2^j + 2^3_i 2_3^j)\epsilon_j\,, \nonumber\\
  \delta'_Q \bar\zeta^{1,2,3} = 0 &\Rightarrow& {\bar\epsilon'}{}^i =
  2^i_4 2^4_j \bar\epsilon^j\,.  \label{5.102}
\end{eqnarray}
Now, in the analytic basis at point 1 (the analog of (\ref{3.9})) $x^\mu$
transforms as follows:
\begin{equation}\label{5.103}
  \delta_Q x^\mu = -2i[\theta_2\sigma^\mu\bar\epsilon^i 1^2_i +
  \theta_3\sigma^\mu\bar\epsilon^i 1^3_i +
  \theta_4\sigma^\mu\bar\epsilon^i 1^4_i- 1^i_1 \epsilon_i \sigma^\mu
  \bar\theta^1]\,.
\end{equation}
Replacing the parameters in (\ref{5.103}) by the residual ones from
(\ref{5.102}) we can find $\delta'_Q x^\mu$. Then it is easy to verify that the
following combination
\begin{equation}\label{5.104}
  x^\mu- y^\mu + \frac{2i}{(1^1 2_4)}  \xi_{2/4}\sigma^\mu \bar\theta^1\,, \qquad \xi_{2/4}
  \equiv [(1^2 2_4) \theta_2 + (1^3 2_4) \theta_3 +
  (1^4 2_4) \theta_4]
\end{equation}
is invariant under the residual $Q$ supersymmetry (note that $\delta'_Q
y^\mu=0$). This allows us to write the two-point function (\ref{5.101}) in the
frame $\zeta=0$ in the form of a coordinate shift of the leading component
(\ref{5.3}):
\begin{eqnarray}
  \langle W^{1}(\theta)
  \bar W_4(0) \rangle
  &=& \exp\left\{ \frac{2i}{(1^1 2_4)} \xi_{2/4} \sigma^\mu \bar\theta^1
  \frac{\partial}{\partial x^\mu} \right\}
  \frac{(1^{1}2_4)}{
  (x-y)^2} \label{5.105} \\
  &=& \left\{ 1 + \frac{2i}{(1^1 2_4)} \xi_{2/4}
  \sigma^\mu \bar\theta^1 \partial_{x\; \mu} - \frac{1}{(1^1 2_4)^2} (\xi_{2/4})^2 (\bar\theta^1)^2 \square_x \right\}
  \frac{(1^{1}2_4)}{
  (x-y)^2}\,. \nonumber
\end{eqnarray}
Now it becomes clear that the coordinate shift can be the source
of harmonic singularities of the type $(1^1 2_4)^{-1}$ or $(1^1
2_4)^{-2}$, since $(1^1 2_4) = 0$ when points 1 and 2 coincide.
However, this does not happen in (\ref{5.105}) because the leading
component (\ref{5.3}) already contains one factor $(1^1 2_4)$ and
because $\square_x (x-y)^{-2} = 0$ if $x^\mu \neq y^\mu$. So, we
can say that the two-point function (\ref{5.101}) is free from
harmonic singularities at point 1 (and similarly at point 2), at
least up to space-time contact terms. This can also be translated
into the statement that the full superfunction (\ref{5.101})
satisfies the H-analyticity condition (\ref{5.5}) with
covariantized harmonic derivatives (cf. (\ref{covdev})).

\subsection{Three-point functions in the case ${\cal O}^{[m,0,0]}_{1/2}
  {\cal O}^{[0,0,n]}_{1/2}$}

The discussion of the two-point function above can easily be adapted to the
three-point function (\ref{5.1'}). Firstly, the $SU(4)$ irrep carried by ${\cal
O}(3)$ should be in the decomposition of the tensor product of the two irreps
at points 1 and 2:
\begin{equation}\label{5.2}
  [m,0,0]\otimes [0,0,n] = \bigoplus_{k=0}^{\min(m,n)}
  [m-k,0,n-k]\,.
\end{equation}
{}From (\ref{0.3}) we see that irreps of the type $[p,0,q]$ can be
realized as products of $W^1$'s and $\bar W^4$'s. This suggests to
build up the $SU(4)$ structure of the function (\ref{5.1'}) in the
form of a product of two-point functions of $W$'s.

Secondly, in addition to the $SU(4)$ quantum numbers, the operator
${\cal O}(3)$ also carries spin and dimension. Since the leading
components at points 1 and 2 are scalars, the Lorentz irrep at
point 3 must be of the type $(s/2,s/2)$, i.e., a symmetric
traceless tensor of rank $s$. The corresponding conformal tensor
structure is built out of the conformally covariant vector
\begin{equation}\label{5.8}
  Y^\mu = \frac{(x-z)^\mu}{(x-z)^2} - \frac{(y-z)^\mu}{(y-z)^2}
\end{equation}
where $z^\mu$ is the space-times coordinate at point 3. Thus, to
the $PU(2,2/4)$ UIR ${\cal D}$ at point 3 corresponds the
following leading term:
\begin{equation}\label{5.9}
  \langle {\cal O}^{[m,0,0]}_{1/2}
  {\cal O}^{[0,0,n]}_{1/2}
  {\cal O}^{(\ell;\ s/2,s/2;\ m-k,0,n-k)}\rangle_{0} =
\end{equation}
$$\left[\frac{(1^1 2_4)}{(x-y)^2}\right]^k
  \left[\frac{(1^1 3_4)}{(x-z)^2}\right]^{m-k}
   \left[\frac{(3^1 2_4)}{(z-y)^2}\right]^{n-k}
   \ (Y^2)^{\frac{1}{2}(\ell-s-m-n+2k)}\ Y^{\{\mu_1}\cdots
  Y^{\mu_s\}}  $$
where $\{\mu_1\cdots\mu_s\}$ denotes traceless symmetrization.

Thirdly, to study the harmonic singularities at point 1 we need
only restore the $\theta$ dependence at this point. This can be
done by analogy with the two-point function above. We start by
fixing a frame in which the odd coordinates are set to zero at
points 2 and 3. This time, in order to reach the frame we have to
use both $Q$ and $S$ supersymmetry. It is important that the
harmonics $1^I_i$ do not participate in the frame fixing, thus
there is no danger of creating artificial harmonic singularities
at point 1. The residual $Q+S$ supersymmetry involves as many
parameters as the remaining number of $\theta$'s at point 1. Next,
the vector $Y^\mu$ (\ref{5.8}) is invariant under conformal boosts
at points 1 and 2 and covariant at point 3 (it is translation
invariant as well). So, we need to find a superextension of
$Y^\mu$ which is invariant under $Q+S$ supersymmetry at points 1
and 2. Remarkably, the combination (\ref{5.104}) that was $Q$
invariant in the two-point case turns out to be $Q+S$ invariant in
this new frame. Thus, performing the shift (\ref{5.104}) on the
vectors $Y$ in (\ref{5.9}) we obtain the desired superextension.
In addition, the two-point factor $(1^1 2_4)/(x-y)^2$ in
(\ref{5.9}) undergoes the same shift, whereas the shift of the
factor $(1^1 3_4)/(x-z)^2$ is obtained from (\ref{5.104}) by
replacing the harmonics $2^I_i$ by $3^I_i$ (the factor $(3^1 2_4)/
(z-y)^2$ requires no supersymmetrization in this frame).

In this way we arrive at the following form of the three-point function where
the full $\theta$ dependence at point 1 is restored:
\begin{eqnarray}
  &&\langle {\cal O}^{[m,0,0]}_{1/2}(\theta)
  {\cal O}^{[0,0,n]}_{1/2}(0)
  {\cal O}^{(\ell;\ s/2,s/2;\ m-k,0,n-k)}(0)\rangle \label{5.10}\\
  &&= \left[\frac{(3^1 2_4)}{(z-y)^2}\right]^{n-k}\times
  \exp\left\{ \frac{2i}{(1^1 3_4)} \xi_{3/4} \sigma^\mu \bar\theta^1
  {\partial_{x\; \mu}} \right\} \left[\frac{(1^1 3_4)}{(x-z)^2}\right]^{m-k} \nonumber\\
  &&\quad \times \exp\left\{ \frac{2i}{(1^1 2_4)} \xi_{2/4} \sigma^\nu \bar\theta^1
  {\partial_{x\; \nu}} \right\} \left\{\left[\frac{(1^1 2_4)}{(x-y)^2}\right]^k
  \ (Y^2)^{\frac{1}{2}(\ell-s-m-n+2k)}\ Y^{\{\mu_1}\cdots
  Y^{\mu_s\}} \right\} \,.\nonumber
\end{eqnarray}
Let us now turn to the crucial question of harmonic singularities. In
(\ref{5.10}) they may originate from the two shifts. Recalling (\ref{5.105}),
we see that no singularity arises when $(1^1 3_4)=0$ (up to space-time contact
terms), just like in the two-point function. However, the presence of a
singularity when $(1^1 2_4)=0$ depends on the value of $k$ and there are three
distinct cases.

{\it (i)}  If $k=0$ the singularity occurs in the $\theta\bar\theta$ term. In
order to remove it we must require:
\begin{equation}\label{5.14}
  \partial^\nu_x \left\{(Y^2)^{\frac{1}{2}(\ell-s-m-n)}\ Y^{\{\mu_1}\cdots
  Y^{\mu_s\}} \right\} = 0
\end{equation}
which implies
\begin{equation}\label{5.15}
  s=0\,, \quad \ell = a_1+a_2+a_3
\end{equation}
where $[a_1,a_2,a_3]$ is the $SU(4)$ irrep at point 3 (in the case
at hand $a_1=m,a_2=0,a_3=n$). In this case we can immediately
write down the complete three-point function (\ref{5.1'}) in the
form of a product of two-point functions of the type
(\ref{5.101}):
\begin{equation}\label{5.16}
  \langle {\cal O}^{[m,0,0]}_{1/2}(1)
  {\cal O}^{[0,0,n]}_{1/2}(2)
  {\cal O}^{(m+n;\ 0,0;\ m,0,n)}(3)\rangle =
  \langle W^1(1)\bar W_4(3) \rangle^{m}
  \ \langle W^1(3)\bar W_4(2) \rangle^{n} \,.
\end{equation}
Consequently, ${\cal O}^{(m+n;\ 0,0;\ m,0,n)}$ is a 1/4 BPS short
{\it protected operator}.

{\it (ii)} If $k=1$ the singularity occurs in the $(\theta\bar\theta)^2$ term.
In order to remove it we must require:
\begin{equation}\label{5.11}
  \square_x \left\{(x-y)^{-2}
  \ (Y^2)^{\frac{1}{2}(\ell-s-m-n+2)}\ Y^{\{\mu_1}\cdots
  Y^{\mu_s\}} \right\} = 0\,.
\end{equation}
This equation admits two solutions: $\ell = s + m +n$, i.e.,
\begin{equation}\label{5.12}
  \mbox{\it (ii.a)} \qquad \ell  = 2+s+a_1+a_2+a_3\,.
\end{equation}
Note that this corresponds to the unitarity bound
(\ref{unitbound}). From Section 2 we know that this is a necessary
condition for ${\cal O}^{\cal D}(3)$ to be a current-like object.
Remarkably, it turns out to be sufficient as well, as can be seen
by restoring the $\theta$ dependence only at point 3 and then
checking that the constraints (\ref{currcon1}) or (\ref{currcon2})
are automatically satisfied (the details of this calculation will
be given elsewhere). Thus, we conclude that ${\cal O}^{(s+m+n;\
s/2,s/2;\ m-1,0,n-1)}$ - being forced to saturate the unitarity bound -
 is a current-like {\it protected operator}.

The second solution is
$$
   \ell = -s+a_1+a_2+a_3\,.
$$
This solution clearly violates the unitarity bound for series A
(\ref{seriesA}). It can only be compatible with the series C if we
set $s=0$:
\begin{equation}\label{5.13}
  \mbox{\it (ii.b)} \qquad s=0\,, \quad \ell = a_1+a_2+a_3\,.
\end{equation}
Once again, the three-point function is reduced to a product of two-point
functions:
\begin{eqnarray}
  &&\langle {\cal O}^{[m,0,0]}_{1/2}(1)
  {\cal O}^{[0,0,n]}_{1/2}(2)
  {\cal O}^{(m+n-2;\ 0,0;\ m-1,0,n-1)}(3)\rangle \label{5.12'}\\
  &&\qquad = \langle W^1(1)\bar W_4(2)
  \rangle\ \langle W^1(1)\bar W_4(3) \rangle^{m-1}
  \ \langle W^1(3)\bar W_4(2) \rangle^{n-1}\,.  \nonumber
\end{eqnarray}
{}From this form it is clear that ${\cal O}^{(m+n-2;\ 0,0;\ m-1,0,n-1)}$ is a
1/4 BPS short {\it protected operator} of the type (\ref{0.3}). Exceptionally,
if $\min(m,n)=1$, the operator becomes 1/2 BPS short.

{\it (iii)} If $k\geq 2$ the harmonic factor $(1^1 2_4)^k$ is
sufficient to suppress all the singularities originating from the
coordinate shift. Then the operator ${\cal O}(3)$ can either
belong to series C (if $s=0$ and $\ell=m+n-2k$) or to series A. In
the latter case its dimension is only limited by the unitarity
bound (\ref{seriesA}), so in general it is an {\it unprotected
operator}. If the unitarity bound happens to be saturated, $\ell =
2+s+m+n-2k$, the operator ${\cal O}(3)$ becomes current-like.

The results of this subsection are summarized in Table 1.

\subsection{The case ${\cal O}^{[0,m,0]}_{1/2}
  {\cal O}^{[0,n,0]}_{1/2}$}

The three-point function (\ref{5.1}) can be treated in very much the same way.
Leaving the details for a future publication, here we only mention a few key
points.

The analog of the two-point function (\ref{5.101}) now is (leading component
only)
\begin{equation}\label{5.3'}
  \langle W^{12}(x,0,1) W^{12}(y,0,2)\rangle = \frac{(1^{12}2^{12})}{
  (x-y)^2}
\end{equation}
where
\begin{equation}\label{5.4'}
  (1^{12}2^{12}) = \epsilon^{ijkl}1^1_i1^2_j2^1_k2^2_l\,.
\end{equation}
The $\theta$ dependence at point 1 can be restored by making the shift (cf.
(\ref{5.104}))
\begin{equation}\label{5.104'}
  x^\mu- y^\mu - \frac{2i}{(1^{12}2^{12})} [\xi^{2/12}\sigma^\mu \bar\theta^1
  -\xi^{1/12}\sigma^\mu \bar\theta^2] \,, \qquad \xi^{I/12}
  \equiv [(1^{I3}2^{12}) \theta_3 + (1^{I4}2^{12}) \theta_4]\,.
\end{equation}
The corresponding exponential (cf. (\ref{5.105})) now contains terms up to the
level $(\theta\bar\theta)^4$, but careful examination shows that in it one does
not encounter singularities worse than $(1^{12}2^{12})^{-2}$.

The $SU(4)$ irrep carried by ${\cal O}(3)$ should be in the decomposition of
the tensor product (we assume that $m\geq n$):
\begin{equation}\label{5.2'}
  [0,m,0]\otimes [0,n,0] = \bigoplus_{k=0}^n
  \bigoplus_{j=0}^{n-k}[j,m+n-2j-2k,j]\,.
\end{equation}
{}From (\ref{4.163}) we see that irreps of the type $[q,p,q]$ can
be obtained as products of the two G-analytic realizations of the
$N=4$ SYM field strength (\ref{4.16}) and (\ref{4.160}). So, we
also need the mixed two-point function
\begin{equation}\label{5.6'}
  \langle W^{12}(x,0,u_1) W^{13}(y,0,u_2)\rangle = \frac{(1^{12}2^{13})}{
  (x-y)^2}\,.
\end{equation}
Then, generalizing (\ref{5.9}), we can write down the leading
term\footnote{Note that if $m=n$ and the two operators at points 1 and 2 are
considered identical, some of the terms in (\ref{5.2'}) drop out due to
crossing symmetry.}
\begin{eqnarray}
&& \langle {\cal O}^{[0,m,0]}_{1/2}(1)
  {\cal O}^{[0,n,0]}_{1/2}(2)
  {\cal O}^{(\ell;\ s/2,s/2;\ j,m+n-2j-2k,j)}(3)\rangle_{0} = \nonumber\\
 &&\quad \left[\frac{(1^{12}2^{12})}{(x-y)^2}\right]^k
  \left[\frac{(1^{12}3^{12})}{(x-z)^2}\right]^{m-j-k}
   \left[\frac{(2^{12}3^{12})}{(y-z)^2}\right]^{n-j-k} \nonumber\\
  &&\quad \times \left\{\left[\frac{(1^{12}3^{12})}{(x-z)^2}\right]
   \left[\frac{(2^{12}3^{13})}{(y-z)^2}\right]
  - \left[\frac{(1^{12}3^{13})}{(x-z)^2}\right]
   \left[\frac{(2^{12}3^{12})}{(y-z)^2}\right]\right\}^j  \nonumber\\
  &&\quad \times\ (Y^2)^{\frac{1}{2}(\ell-s-m-n+2k)}\ Y^{\{\mu_1}\cdots
  Y^{\mu_s\}} \,. \label{5.8'}
\end{eqnarray}

Finally, we restore the $\theta$ dependence at point 1 by making shifts of the
type (\ref{5.104'}) and study the resulting harmonic singularities in the
variable $(1^{12}2^{12})$. They depend on the value of $k$ and we find exactly
the same conditions (\ref{5.15}), (\ref{5.12}) and (\ref{5.13}) as in the
preceding case (see the summary in Table 2).

\newpage

\begin{table}[h]
  \begin{center}
    \leavevmode
\caption{${\cal O}^{[m,0,0]}_{1/2}\; {\cal O}^{[0,0,n]}_{1/2} \ \rightarrow \
  \sum_{k=0}^{\min(m,n)}
  {\cal O}^{(\ell;s/2,s/2;m-k,0,n-k)}$ } \label{tab1}
    \begin{tabular}{lllll}
  \\ \\
\hline
 k & Spin  & Dimension & Protection & Type \\ \hline
  \\
 $k=0$ & $s=0$ & $\ell=m+n$ & protected & 1/4 BPS \\
  \\
 $k=1$ & $s=0$ & $\ell=m+n-2$ & protected & 1/4 BPS \\
       &       &              &           & 1/2 BPS if $\min(m,n)=1$ \\
               & $s\geq0$ & $\ell=s+m+n$  & protected & current-like \\
  \\
 $k\geq 2$ & $s=0$ & $\ell = m+n-2k$  & protected & 1/4 BPS \\
  &       &              &           & 1/2 BPS if $\min(m,n)=k$ \\
        & $s\geq0$ & $\ell = 2+s+m+n-2k$  & ``protected'' & current-like \\
        &    & $\ell > 2+s+m+n-2k$  & unprotected & long \\     \\
    \end{tabular}
  \end{center}

  \begin{center}
    \leavevmode
\caption{${\cal O}^{[0,m,0]}_{1/2}\; {\cal O}^{[0,n,0]}_{1/2} \ \rightarrow \
  \sum_{k=0}^{n}\sum_{j=0}^{n-k}
  {\cal O}^{(\ell;s/2,s/2;j,m+n-2j-2k,j)}\,,\qquad m\geq n$ } \label{tab2}
    \begin{tabular}{lllll}
  \\ \\
\hline
 k & Spin  & Dimension & Protection & Type \\ \hline
  \\
 $k=0$ & $s=0$ & $\ell=m+n$ & protected & 1/4 BPS \\
       &       &            &           & 1/2 BPS if $j=0$\\
  \\
 $k=1$ & $s=0$ & $\ell=m+n-2$ & protected & 1/4 BPS \\
        &       &            &            & 1/2 BPS if $j=0$\\
     & $s\geq0$ & $\ell=s+m+n$  & protected & current-like \\
  \\
 $k\geq 2$ & $s=0$ & $\ell = m+n-2k$  & protected & 1/4 BPS \\
  &       &              &           & 1/2 BPS if $j=0$ \\
        & $s\geq0$ & $\ell = 2+s+m+n-2k$  & ``protected" & current-like \\
        &    & $\ell > 2+s+m+n-2k$  & unprotected & long \\
    \end{tabular}
  \end{center}
\end{table}

\section{Conclusions}

In this paper we have established the OPE content of two types of 1/2 BPS short
operators. The results are summarized in Tables \ref{tab1} and \ref{tab2}.  The
most interesting case is that of $N=4$ SYM (Table \ref{tab2}), where the
central charge vanishes and we are actually dealing with the superalgebra
$PSU(2,2/4)$.

Let us give two examples of operators widely discussed in the literature. They
appear in the OPE of two $N=4$ SYM stress-tensor multiplets ($m=n=2$).

The case $s=0$, $\ell=4$, $k=1$ and $SU(4)$ irrep $[0,2,0]$ corresponds to the
so-called ${\cal O}^4_{20'}$ whose non-renormalization was first conjectured in
\cite{AFP1}. As we can see from Table \ref{tab2}, it is a  {\it protected
current-like} operator because the OPE fixes its dimension.

The case $s=0$, $\ell\geq 2$, $k=2$  and $SU(4)$ irrep $[0,0,0]$ corresponds to
the so-called Konishi multiplet \cite{Konishi}.  This time the OPE does not fix
the dimension, therefore the operator is either current-like if $\ell=2$ (which
only takes place in the free theory) or long if $\ell>2$ (in the presence of
interactions this operator picks an anomalous dimension).

We stress that in the OPE of two $N=4$ SYM stress-tensor multiplets the only
unprotected operators are $SU(4)$ singlets (since $m=n=k=2$ implies $j=0$,
i.e., Dynkin labels $[0,0,0]$). As an application of this result, consider the
correlator of four stress-tensor multiplets
\begin{equation}
G^{(N=4)} = \langle O^{[0,2,0]}(1) \, O^{[0,2,0]}(2) \, O^{[0,2,0]}(3) \,
O^{[0,2,0]}(4) \rangle \, .
\end{equation}
Performing a double OPE, we see that the six $SU(4)$ representations
$[j,4-2k-2j,j]$, with $k=0,1,2$ and $0 \leq j \leq 2-k$ can be exchanged.
According to our classification, the only ``unprotected channel'' is the
singlet $j=0, k=2$ which is just the operator  ${\cal O}^4_{20'}$ mentioned
above.

As another application of our results, we sketch the explanation of the
non-renormalization of the so-called ``extremal" and ``next-to-extremal"
$n$-point correlators \cite{DHoFrMaMaRa,EdHoScSoWe}.

Consider an ``extremal'' 4-point function of operators
$O^{[0,m,0]} = [W^{12}]^m$, where the exponents (viz charges) obey
$m_1 = m_2 + m_3 + m_4$.

In the OPE between the operators at points 1 and 2 we find a range
of representations with Dynkin labels $[j,m_1+m_2-2j -2k,j]$.
Since $m_1>m_2$ we must have $k \leq m_2$. The number of indices
of the corresponding Young tableau is $2(m_1 + m_2 - 2k)$. The
representation with the smallest number of boxes in its tableau
 therefore has
\begin{equation}
2 (m_1 + m_2 - 2 m_2) = 2 (m_3 + m_4)
\end{equation}
indices. This equals the total number of indices of the remaining
operators at points 3 and 4. The only operators in their OPE
contributing to the correlator carry therefore SU(4) Dynkin labels
$[j,m_i+m_j-2j,j]$, i.e. the case $k=0$ in Table 2, because $k>0$
means a dualization by the $\epsilon$ symbol. Only 1/4 BPS or 1/2
BPS operators are exchanged, and they always have protected
dimension.

For the ``next-to-extremal'' case $m_1 = m_2 + m_3 + m_4 - 2$ the
same argument leads to one $\epsilon$-dualization in the product
of the operators at points 3,4. Hence, the exchanged operators
carry Dynkin labels $[j,m_i+m_j-2j-2,j]$ corresponding to the case
$k=1$ in Table 2. Again, all these operators are protected.

\section*{Acknowledgements}
At the early stages of this work we profited a lot from
discussions with G. Arutyunov and A. Petkou. We are grateful to S.
Ferrara for his constructive critical remarks.



\begin{thebibliography}{99}

\bibitem{FGG} S. Ferrara, R. Gatto and A.F. Grillo, \npb34(1971)349;
\ap76(1973)161; S. Ferrara, R. Gatto, A.F. Grillo and G. Parisi,
\npb49(1972)77.

\bibitem{DPPT} V.K. Dobrev, V.B. Petkova, S.G. Petrova and I.T.Todorov,
\prd13(1976)887.

\bibitem{Pisa} I.T. Todorov, M.C. Mintchev and V.B. Petkova,
{\em Conformal invariance in quantum field theory}, Pisa, Italy:
Sc. Norm. Sup. (1978) 273p.

\bibitem{FP}
E.S. Fradkin and M.Ya. Palchik, \rep44(1978)249; {\em Conformal
Quantum Field Theory in D-Dimensions}, Kluwer (1996) 461p.

\bibitem{FGPa}
S. Ferrara, A.F. Grillo and G. Parisi,
Lett. Nuovo Cim. \textbf{5} (1972) 147.

\bibitem{Anselmi:1997mq}
D. Anselmi, M. Grisaru and A.A. Johansen,
Nucl. Phys. \textbf{B491} (1997) 221, \xxx9601023;
D. Anselmi, D.Z. Freedman, M.T. Grisaru and A.A. Johansen,
Phys. Lett. \textbf{B394} (1997) 329, \xxx9608125;
E. D'Hoker, S.D. Mathur, A. Matusis and L. Rastelli,
Nucl.  Phys. \textbf{B589} (2000) 38, \xxx9911222;
L. Hoffmann, A.C. Petkou and W. Ruehl,
{\em Aspects of the conformal operator product expansion in AdS/CFT
correspondence}, \xxx0002154;
M. Bianchi, S. Kovacs, G. Rossi and Y.S. Stanev,
JHEP {\bf 9908} (1999) 020, \xxx9906188.

\bibitem{HWOPE}
P.S. Howe and P.C. West, \plb389(1996)273, \xxx9607060.

\bibitem{hh}
G.G. Hartwell and P.S. Howe, Int. J. Mod. Phys. \textbf{A10} (1995)
3901, \xxx9412147; Class. Quant. Grav. \textbf{12} (1995) 1823.

\bibitem{Pickering:2000rk}
A. Pickering and P.C. West,
Nucl. Phys. \textbf{B569} (2000) 303, \xxx9904076.

\bibitem{DolOsb}
F.A. Dolan and H. Osborn,
\npb593(2001)599, \xxx0006098.


\bibitem{GIO}
A. Galperin, E. Ivanov and V.I. Ogievetsky,
JETP Lett. \textbf{33} (1981) 168.

\bibitem{GIK1}
A. Galperin, E. Ivanov, S. Kalitzin, V. Ogievetsky and E. Sokatchev,
Class. Quant. Grav. \textbf{1} (1984) 469.

\bibitem{GIK2}
A. Galperin, E. Ivanov, S. Kalitzin, V. Ogievetsky and E. Sokatchev,
Class. Quant. Grav. \textbf{2} (1985) 155;
R. Kallosh, Pis'ma ZHETF \textbf{41} (1985) 172;
A.A. Rosly and A.S. Schwarz,
Commun. Math. Phys. \textbf{105} (1986) 645.

\bibitem{Bandos}
I. Bandos,
Theor. Math. Phys. \textbf{76} (1988) 783.

\bibitem{dp}
V. K. Dobrev and V. B. Petkova,
Phys. Lett. \textbf{B162} (1985) 127;
Fortschr. Phys. \textbf{35} (1987) 7, 537.

\bibitem{Gunaydin:1985fk}
M. Gunaydin and N. Marcus,
Class. Quant. Grav. {\bf 2} (1985) L11.

\bibitem{AF}
L. Andrianopoli and S. Ferrara, Phys. Lett. \textbf{B430} (1998)
248, \xxx9803171; Lett. Math. Phys. \textbf{46} (1998) 265,
\xxx9807150; Lett. Math. Phys. \textbf{48} (1999) 145,
\xxx9812067; S. Ferrara, M. Porrati and A. Zaffaroni,
Lett. Math. Phys. \textbf{47} (1999) 255.

\bibitem{Arutyunov:2001qw}
G. Arutyunov, B. Eden and E. Sokatchev,
{\em On non-renormalization and OPE in superconformal field theories},
\xxx0105254.


\bibitem{AFP1}
G. Arutyunov, S. Frolov and A.C. Petkou,
\npb586(2000)547, \xxx0005182.

\bibitem{AFP2}
G. Arutyunov, S. Frolov and A.C. Petkou, \npb602(2001)238,
\xxx0010137.

\bibitem{AEPS}
G. Arutyunov, B. Eden, A.C. Petkou and E. Sokatchev, {\em
Exceptional non-renormalization properties and OPE analysis of
chiral  four-point functions in N = 4 SYM(4)}, \xxx0103230.

\bibitem{BKRSkonishi}
M. Bianchi, S. Kovacs, G. Rossi and Y.S. Stanev,
\jhep0105(2001)042, \xxx0104016.

\bibitem{TheBook}
A. Galperin, E. Ivanov, V. Ogievetsky and E. Sokatchev, {\em Harmonic
superspace}, CUP, to appear.

\bibitem{Park}
J. Park,
\ijmp13(1998)1743, \xxx9703191.

\bibitem{Osborn}
H. Osborn,
\ap272(1999)243, \xxx9808041.

\bibitem{Park2}
J. Park,
Nucl. Phys. {\bf B559} (1999) 455, \xxx9903230.

\bibitem{hw9}
P.S. Howe and P.C. West,
{\em AdS/SCFT in superspace}, \xxx0105218.

\bibitem{hw1}
P.S. Howe and P.C. West, \ijmp14(1999)2659, \xxx9509140;
\plb400(1997)307, \xxx9611075; {\em Is $N=4$ Yang-Mills soluble?}
(1996) in Moscow 1996, \xxx9611074.

\bibitem{EHW}
B. Eden, P.S. Howe and P.C. West, \plb463(1999)19, \xxx9905085.

\bibitem{HoSoWe}
P.S. Howe, E. Sokatchev and P.C. West,
\plb444(1998)341, \xxx9808162.

\bibitem{KuzTheis}
S.M. Kuzenko and S. Theisen,
Class. Quant. Grav. {\bf 17} (2000) 665, \xxx9907107.

\bibitem{DHoFrMaMaRa}
E. D'Hoker, D.Z. Freedman, S.D. Mathur, A. Matusis and L. Rastelli,
{\em Extremal correlators in the AdS/CFT correspondence},
\xxx9908160.

\bibitem{EdHoScSoWe}
B.U. Eden, P.S. Howe, C. Schubert, E. Sokatchev and P.C. West,
\plb472(2000)323, \xxx9910150;

\bibitem{Pickering}
B.U. Eden, P.S. Howe, A. Pickering, E. Sokatchev and P.C. West,
\npb581(2000)523, \xxx0001138.

\bibitem{Rosly} A.A. Rosly,
{\em Constraints in supersymmetric Yang-Mills theory as
integrability conditions}, in {\em Theor. Group Methods in Physics}
Editors M. Markov, V. Man'ko, Nauka, Moskwa, (1983) 263;
Class. Quantum. Grav. \textbf{2} (1985) 693.

\bibitem{Siegel}
W. Siegel, Nucl. Phys. \textbf{B177} (1981) 325;
P.S. Howe, K.S. Stelle and P.K. Townsend,
Nucl. Phys. \textbf{B191} (1981) 445; Nucl. Phys. \textbf{B192} (1981) 332.

\bibitem{AFSZ}
L. Andrianopoli, S. Ferrara, E. Sokatchev and B. Zupnik,
Adv. Theor. Math. Phys. \textbf{3} (1999) 1149, \xxx9912007.

\bibitem{Ferrara:2000zg}
S. Ferrara and E. Sokatchev,
Lett. Math. Phys.  {\bf 52} (2000) 247, \xxx9912168.

\bibitem{Ferrara:2001eb}
S. Ferrara and E. Sokatchev,
Int. J. Theor. Phys. {\bf 40} (2001) 935, \xxx0005151.

\bibitem{Konishi} K. Konishi, Phys. Lett. {\bf B135} (1984) 439.

\end{thebibliography}
\end{document}